\documentclass[preprint,amsmath,amssymb,aps,pre,superscriptaddress]{revtex4-1}

\usepackage{xcolor}
\usepackage{graphicx}
\usepackage{dcolumn}
\usepackage{mathtools}

\begin{document}

\title{The Rayleigh collapse of two spherical bubbles}

\author{Anthony Harkin}
\affiliation{School of Mathematical Sciences, Rochester Institute of Technology, Rochester, NY 14623, USA}
\email{corresponding author:harkin@rit.edu}
\author{Adam Giammarese}
\affiliation{School of Mathematical Sciences, Rochester Institute of Technology, Rochester, NY 14623, USA}
\author{Nathaniel S. Barlow}
\affiliation{School of Mathematical Sciences, Rochester Institute of Technology, Rochester, NY 14623, USA}
\author{Steven J. Weinstein}
\affiliation{Department of Chemical Engineering, Rochester Institute of Technology, Rochester, NY 14623, USA}
\affiliation{School of Mathematical Sciences, Rochester Institute of Technology, Rochester, NY 14623, USA}

\date{\today}     

\begin{abstract}
The inertial collapse of two interacting and non-translating spherical bubbles of equal size is considered. 
The exact analytic solution to the nonlinear ordinary differential equation that governs the bubble radii 
during collapse is first obtained via a slowly converging power series.  An asymptotic approximant is then
constructed that accelerates convergence of the series and imposes the asymptotic collapse behavior when 
the radii are small.  The solution generalizes the classical 1917 Rayleigh problem of single bubble collapse,
as this configuration is recovered when the distance between the bubble centers far exceeds that of their radii.
\end{abstract}

\maketitle

\section{Introduction}

In 1917 Lord Rayleigh examined the following problem:
``An infinite mass of homogeneous  incompressible fluid acted upon by 
no forces is at rest, and a spherical portion of the fluid is suddenly
annihilated; it is required to find the instantaneous alteration of
pressure at any point of the mass, and the time in which the cavity 
will be filled up, the pressure at an infinite distance being supposed 
to remain constant." Rayleigh suggested a 
possible connection between imploding bubbles and the erosion damage found 
on ship propellers~\cite{LR1917}. 
Ignoring surface tension and thermal effects, 
Rayleigh derived an equation that governs the collapse of an empty spherical 
bubble of radius $R(T)$ surrounded by an incompressible inviscid 
liquid of density, $\rho$:
\begin{equation}
R\frac{d^2R}{dT^2}  + \frac{3}{2}\left( \frac{dR}{dT} \right)^2 + \frac{p_{L}}{\rho}
= 0,
\label{Rayleigheqn}
\end{equation}
where $p_{L}$ is the ambient liquid pressure which is assumed constant.
The Rayleigh equation~(\ref{Rayleigheqn}) and its generalizations have been widely used to understand phenomena involving gas bubbles that have enabled many useful technologies~\cite{B1995,BE1966,LK2010,L2003}.

If we multiply (\ref{Rayleigheqn}) by $ 3R^2\dot{R}$, integrate, and impose that 
the initial bubble radius is at rest with radius $R(0)=R_0$ we obtain an equation 
for $\dot{R}^2$; its square root yields
\begin{equation}
\frac{dR}{dT} = - \left( \frac{2p_{L}}{3\rho} \right) ^{1/2} \sqrt{\frac{R_0^3}{R^3}-1}
\label{modifiedRayleigh}
\end{equation}
where a minus sign is chosen to be consistent with bubble collapse.  The well-known \cite{Besant1859,LR1917,OBF2012} expression for the collapse time, $T_{c1}$, may be obtained by integrating~(\ref{modifiedRayleigh}) to obtain
\begin{equation}
T_{c1}=\xi_1R_0\sqrt{\frac{\rho}{p_L}}
\label{Tc1}
\end{equation}
where
\begin{equation}
\xi_{1} = \sqrt{\frac{3}{2}} \int_0^1 \left( r^{-3} -1 \right) ^{-1/2} dr 
= \sqrt{\frac{3\pi}{2}}\frac{\Gamma(5/6)}{\Gamma(1/3)}
\approx 0.91468\ldots
\label{xi1}
\end{equation}

The solutions $R(t)$ of the Rayleigh equation~(\ref{Rayleigheqn}) and its various generalizations are  most often obtained numerically. Although numerical solutions of~(\ref{Rayleigheqn}) are straightforward, analytical solutions of the Rayleigh equations are useful, especially if the equation is to be embedded in larger system models.  To that end, an infinite series solution of the Rayleigh equation is employed in~\cite{OBF2012} to construct a highly accurate approximation in terms 
of the polylogarithm, and the result compares favorably with high-precision cavitation data obtained in microgravity. A mathematical analysis of the convergence of this approximate polylogarithm solution is performed in~\cite{AF2013}. A parametric solution for the collapsing bubble radius is obtained in terms of hypergeometric functions in~\cite{KS2014, MR2016}.

In this paper, we generalize the Rayleigh problem to consider the collapse of two interacting and non-translating spherical bubbles having the same radii.  We develop a power series solution of the two-bubble Rayleigh equation that is slowly convergent over the entire physical domain and is thus an exact solution to the problem. We then employ the method of asymptotic approximants~\cite{AA2017} to accelerate the convergence.  An asymptotic approximant is defined as a closed-form expression 
whose expansion in one region is exact up to a specified order and whose asymptotic equivalence in another region is enforced~\cite{AA2017}. Here, we assure that the approximant matches the exact power series solution as $T$ approaches zero as well as the asymptotic behavior as $T$ approaches the time of bubble collapse.  The desirable feature of asymptotic approximants is their ability to attain uniform accuracy not only in these two regions, but also at all points in-between, as demonstrated thus far for problems in thermodynamics~\cite{BSWK2012,BSWK2015}, astrophysics~\cite{BH2018}, 
fluid dynamics~\cite{AA2017,FS2020}, and epidemiology~\cite{SIR2020, SEIR2020}.

\section{Collapse time of two interacting spherical bubbles}

 Consider two collapsing spherical bubbles of radius 
$R_{1}(T)$ and $R_{2}(T)$ in a liquid of density, $\rho$, and external pressure, $p_L$, whose centers are separated by a fixed distance $D$. The differential equations 
that govern the interaction of two collapsing and non-translating bubbles are 
provided by~\cite{MAPOL1997,HKN2001,D2001} 
\begin{align}
R_{1}\frac{d^2R_{1}}{dT^2}  + \frac{3}{2}\left( \frac{dR_{1}}{dT} \right)^2 
+\frac{1}{D} 
\left[  R_{2}^{2}\frac{d^{2}R_{2}}{dT^2} + 2R_{2}\left( \frac{dR_{2}}{dT} \right)^{2} \right]
+ \frac{p_{L}}{\rho}
& = 0 
\nonumber \\
R_{2}\frac{d^2R_{2}}{dT^2}  + \frac{3}{2}\left( \frac{dR_{2}}{dT} \right)^2 
+\frac{1}{D} 
\left[  R_{1}^{2}\frac{d^{2}R_{1}}{dT^2} + 2R_{1}\left( \frac{dR_{1}}{dT} \right)^{2} \right]
+ \frac{p_{L}}{\rho}
& = 0.
\label{eq:gov}
\end{align}
where terms of $O(1/D^4)$ are neglected.  We restrict attention to the case where the bubbles are of equal size during collapse, i.e. $R_1(T)=R_2(T)=R(T)$, and thus equations~(\ref{eq:gov}) become:
\begin{equation}
R\frac{d^{2}R}{dT^{2}}  + \frac{3}{2}\left( \frac{dR}{dT} \right)^2 
+\frac{1}{D} 
\left[  R^{2}\frac{d^{2}R}{dT^2} + 2R\left( \frac{dR}{dT} \right)^{2} \right]
+ \frac{p_{L}}{\rho}
= 0
\label{twoequalbubbles}
\end{equation}
with initial conditions $R(0)=R_0, \dot{R}(0)=0$.
If we multiply~(\ref{twoequalbubbles}) by $3R^{2}\dot{R}$, integrate, and apply
the initial conditions to evaluate the resulting constant, we obtain
\begin{equation*}
\frac{3}{2} R^{3} \left( \frac{dR}{dT} \right) ^2  
\left[ 1 + \frac{R}{D} \right]
= \frac{p_{L}}{\rho}(R_0^3 - R^3) .
\end{equation*}   
Solving for $\dot{R}$ gives
\begin{equation}
\frac{dR}{dT} = - \left( \frac{2p_{L}}{3\rho} \right) ^{1/2} \sqrt{\frac{R_0^3}{R^3}-1}
\left[ 1 + \frac{R}{D} \right]^{-1/2},
\label{eq:dRdT}
\end{equation}
where the negative sign is chosen to be consistent with collapse.
Integrating~(\ref{eq:dRdT}) and noting that the radius is zero at full collapse, 
the collapse time, $T_{c2}$ is expressed as 
\begin{subequations}
\begin{equation}
T_{c2}=\left( \frac{3\rho}{2p_{L}} \right) ^{1/2} R_{0} \int_0^1 \left( r^{-3} -1 \right) ^{-1/2} 
\left[ 1 + \epsilon r \right]^{1/2} dr
\label{Tc2}
\end{equation}
where $\epsilon = R_{0}/D$.
Consistent with the accuracy of the dimensional equation~(\ref{eq:gov}), $\epsilon<<1$ is implicitly assumed, and thus the binomial expansion may be used:
\begin{equation}
    \left[ 1 + \epsilon r \right]^{1/2}= \sum_{k=0}^{\infty}  {\frac{1}{2} \choose k} (\epsilon r)^{k}=\frac{\sqrt{\pi}}{2}\sum_{k=0}^{\infty}\frac{(\epsilon r)^k}{\Gamma\left(\frac{3}{2}-k\right)\Gamma(k+1)}.
    \label{eq:binomial}
\end{equation}
The summation~(\ref{eq:binomial}) is convergent, and is substituted into~(\ref{Tc2}).  Upon switching the order of integration and summation, integrals arise of the form:
\begin{equation}
    \int_{0}^1 r^k (r^{-3}-1)^{-1/2} dr=\frac{\sqrt{\pi}~\Gamma\left(\frac{k}{3}+\frac{5}{6}\right)}{(k+1)~\Gamma\left(\frac{k}{3}+\frac{1}{3}\right)}
    \label{eq:gamma}
\end{equation}
\end{subequations}
where $k$ is an integer.  The bubble collapse time~(\ref{Tc2}) is simplified by using the expressions~(\ref{eq:binomial}) and~(\ref{eq:gamma}) such that
\begin{subequations}
\begin{equation}
T_{c2}=R_{0} \sqrt{\frac{\rho}{p_{L}}}\left(\xi_1+\epsilon\eta\right)
\end{equation}
where
\begin{equation}
    \eta= \pi \sqrt{\frac{3}{8}}
\sum_{k=1}^{\infty} \epsilon^{k-1} 
\frac{\Gamma \left( \frac{k}{3}+\frac{5}{6} \right)}{(k+1)\Gamma \left( \frac{k}{3}+\frac{1}{3} \right)\Gamma\left(\frac{3}{2}-k\right)\Gamma(k+1) },
\end{equation}
\label{eq:Tc2}
\end{subequations}
and $\xi_{1}$ is given by (\ref{xi1}).
Although~(\ref{eq:Tc2}) is an exact representation of equation~(\ref{Tc2}), we recall that the original differential equation is accurate with neglected terms of $O(\epsilon^4)$.  For consistency, we utilize just the first 4 terms in the expression for $\eta$ as:
\begin{equation}
\eta=\xi_2+\xi_3\epsilon+\xi_4\epsilon^2+O(\epsilon^3),~~\epsilon<<1
\label{eq:TcTrunc}
\end{equation}
%\begin{align}
%T_{c2} & = \left( \frac{3\rho}{2p_{L}} \right) ^{1/2} R_{0} \int_0^1 \left( r^{-3} -1 \right) ^{-1/2} 
%\left[ \sum_{k=0}^{\infty}  {\sfrac{1}{2} \choose k} (\epsilon r)^{k} \right] dr 
%\\
%& = \left( \frac{3\rho}{2p_{L}} \right) ^{1/2} R_{0} 
%\sum_{k=0}^{\infty} {\sfrac{1}{2} \choose k} \epsilon^{k}
%\int_0^1 r^{k} \left( r^{-3} -1 \right) ^{-1/2} dr 
%\\
%& =
%T_{c2}&=\left( \frac{3\rho}{2p_{L}} \right) ^{1/2} \frac{R_{0}\pi}{2} 
%\sum_{k=0}^{\infty} \epsilon^{k} 
%\frac{\Gamma \left( \frac{k}{3}+\frac{5}{6} \right)}{(k+1)\Gamma \left( \frac{k}{3}+\frac{1}{3} \right)\Gamma\left(\frac{3}{2}-k\right)\Gamma(k+1) } \nonumber \\
%& = R_{0} \sqrt{\frac{\rho}{p_{L}}}\left[\xi_1+\xi_2\epsilon+\xi_3\epsilon^2+\xi_4\epsilon^3+O(\epsilon^4)\right]= R_{0} \sqrt{\frac{\rho}{p_{L}}}\left(\xi_1+\epsilon\eta\right), 
%\label{eq:Tc2}
%\end{align}
where
\begin{align}
\xi_{2} = \frac{\sqrt{3\pi}}{4\sqrt{2}}~\frac{\Gamma(7/6)}{\Gamma(2/3)}
\approx 0.37180\ldots,~\xi_3=\frac{-\pi}{16\sqrt{6}}\approx -0.08016\ldots,\nonumber\\\xi_4=\frac{\sqrt{3\pi}}{64\sqrt{2}}~\frac{\Gamma(11/6)}{\Gamma(4/3)}\approx 0.03573\ldots.
\end{align}
%and, more generally, $\eta$ in~(\ref{eq:Tc2}) is given by
%\begin{equation}
%    \eta= \frac{\pi\sqrt{3}}{2^{3/2}} 
%\sum_{k=1}^{\infty} \epsilon^{k-1} 
%\frac{\Gamma \left( \frac{k}{3}+\frac{5}{6} \right)}{(k+1)\Gamma \left( %\frac{k}{3}+\frac{1}{3} \right)\Gamma\left(\frac{3}{2}-k\right)\Gamma(k+1) }.
%\end{equation}
 In generating plots for this paper, $\eta$ is computed using precisely the truncation in~(\ref{eq:TcTrunc}). Note also that it takes longer for two bubbles to collapse than it does for a single bubble, since $\eta>0$ and $T_{c1}$ is given by~(\ref{eq:Tc2}) with $\epsilon=0$. We also recognize that $\epsilon$ may be chosen to be relatively large owing to the fact that neglected terms are of $O(\epsilon^4)$.
%\begin{equation*}
%T_{c2} = T_{c1} 
%\left( 1 + \frac{\xi_{2}}{\xi_{1}} \epsilon+ \cdots \right).
%\end{equation*}

\section{Power Series Solution and Asymptotic Approximant}

\begin{figure}[h!]
\begin{center}
\includegraphics[width=4.25in]{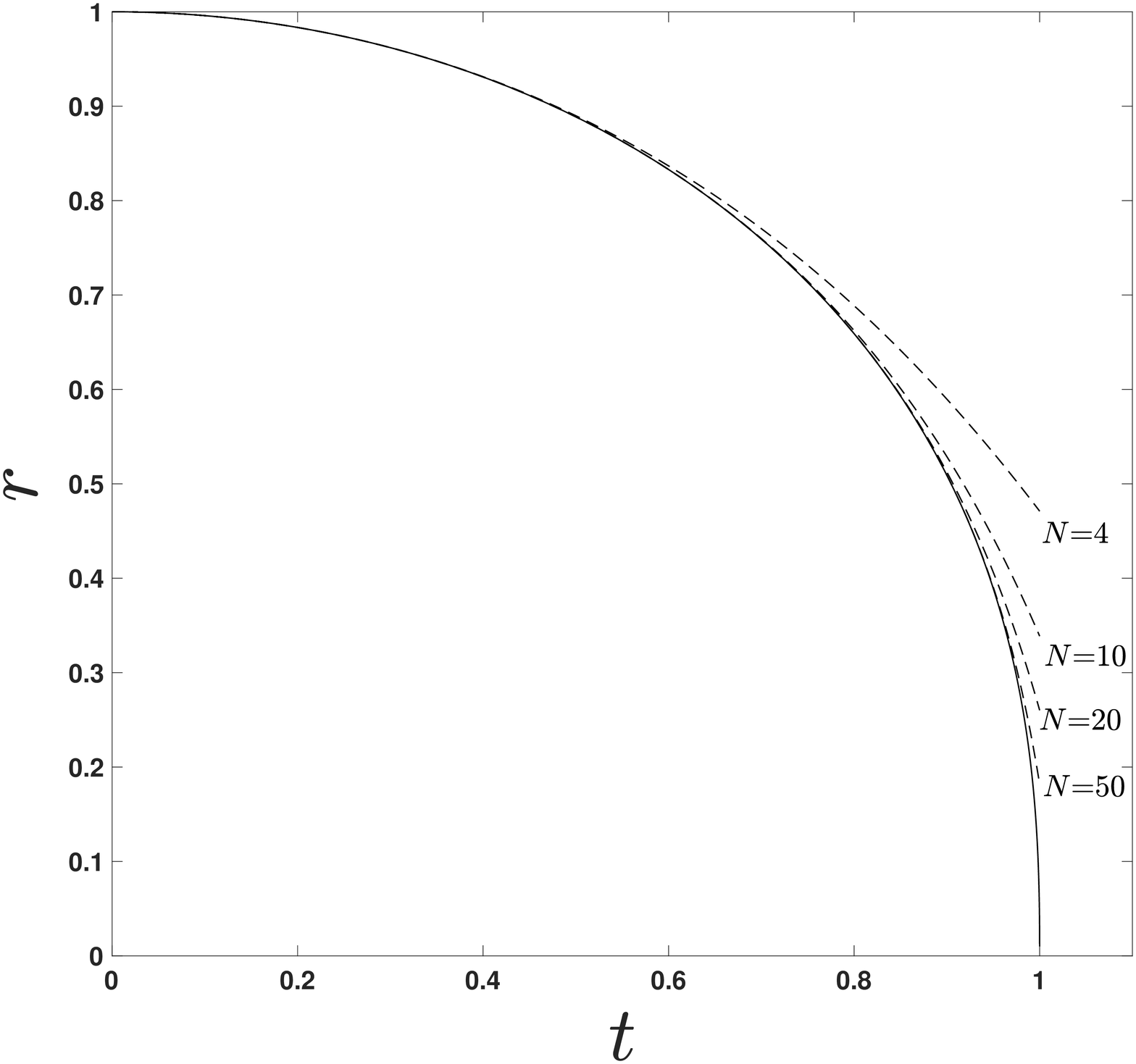}
\end{center}
\vspace*{-0.2in}
\caption{Convergence of the $N$-term partial sums of~(\ref{power_series_two_bubble}) (dashed curve)
of the two-bubble Rayleigh equation~(\ref{rtwobubbles}) ($\epsilon = 0.1$) to the numerical solution (solid curve).}
\label{fig:power_series_two_bubble}
\end{figure}

We nondimensionalize the bubble radii as $R=R_{0}r$, time as $T=T_{c2}t$ and 
express  equation~(\ref{twoequalbubbles}) in dimensionless form as
\begin{equation}
r \frac{d^2r}{dt^2} + \frac{3}{2} \left( \frac{dr}{dt} \right)^{2} 
+\epsilon \left[ r^{2}\frac{d^{2}r}{dt^{2}} + 
2r\left( \frac{dr}{dt} \right)^{2}
\right]
+ \beta=0,
\label{rtwobubbles}
\end{equation}
where $\beta=(\xi_1+\epsilon\eta)^2$ and the initial conditions
are $r(0)=1, \dot{r}(0)=0$.
Note that the two-bubble Rayleigh equation reduces
to the single bubble Rayleigh equation when $\epsilon=0$.
A power series solution around $t=0$ is expressed as
\begin{subequations}
\begin{equation}
r = \sum_{n=0}^{\infty} a_{n} t^{n}.
\label{series}
\end{equation}
where the coefficients $a_n$ are obtained by substitution of~(\ref{series}) and 
its derivatives into~(\ref{rtwobubbles}) yielding
$a_0=1, a_1=0, a_2=-\beta/(2+2\epsilon)$ and
\begin{align}
\nonumber
a_{n+2}=&-\sum_{j=0}^{n-1}\Biggl[a_{j+2}(j+1)(j+2)\left(a_{n-j}+\epsilon\sum_{k=0}^{n-j}a_k a_{n-j-k}\right)\nonumber \\
&+\left(2\epsilon+3/2\right)(j+1)(n-j+1)a_{j+1}a_{n-j+1} \nonumber\\ &+2\epsilon a_{n-j}\sum_{k=0}^j(k+1)(j-k+1)a_{k+1}a_{j-k+1}\Biggr]/\left[(n+2)(n+1)(\epsilon+1)\right], ~~n>0.
\label{a2}
\end{align}
\label{power_series_two_bubble}
\end{subequations}
%a_{n+2}=-\frac{c_n\left(\frac{3}{2}b_0+2\epsilon\right)+b_n\left(\xi_1^2+2\epsilon\xi_1\xi_2\right)
%+\sum_{j=0}^{n-1}\left[\frac{3}{2}c_j %b_{n-j}+\epsilon(j+1)(j+2)a_{j+2}a_{n-j}\right]}{(n+2)(n+1)(\epsilon+1)},
%where
%\begin{equation}
%b_{n>0}=-\sum_{j=1}^n a_j b_{n-j},~b_0=1
%\end{equation}
%and
%\begin{equation}
%c_{n>2}=\frac{1}{2(n-2)a_2}\sum_{j=1}^{n-2}\left(3j-n+2\right)(j+2)a_{j+2}c_{n-j},~c_0=c_1=0,~c_2=4a_2^2.
%\end{equation}
Using the recursion in~(\ref{a2}), the first 4 terms of the power series solution for two collapsing bubbles are 
\begin{equation*}
r(t) = 1
-\frac{\beta}{2\epsilon+2}t^{2}
-\frac{\beta^2(3\epsilon+2)}{12(\epsilon+1)^3}t^{4}
-\frac{\beta^3(75\epsilon^2+102\epsilon+38)}{360(\epsilon+1)^5}t^{6}
- \cdots
\end{equation*}
and this even pattern persists for all orders. The evenness of~(\ref{power_series_two_bubble}) is expected, as~(\ref{rtwobubbles}) is invariant if the independent variable, $t$, is replaced with $-t$~\cite{OBF2012}.  Figure~\ref{fig:power_series_two_bubble} compares the numerical solution of~(\ref{rtwobubbles}) with the power series solution~(\ref{power_series_two_bubble}) for $\epsilon=0.1$.  Note that the numerical solution was obtained via a 4$^\mathrm{th}$ order-accurate Runga-Kutta scheme with $\Delta t=10^{-5}$.  As can be observed, the power series is slowly convergent, and converges non-uniformly, requiring additional terms to preserve accuracy as $t\to1$.  Thus, the power series is impractical to capture the final stages of the bubble collapse.  Note that, although we chose to show the solution to~(\ref{rtwobubbles}) for  $\epsilon=0.1$ here, the solution is nearly indistinguishable for $\epsilon\in[0, 0.5]$ on the scale of figure~\ref{fig:power_series_two_bubble}.

We now develop the asymptotic behavior of the collapse \--- precisely the region that the power series does not well-capture as 
$t \rightarrow 1$
 \-- as follows.  First, we multiply~(\ref{rtwobubbles}a) 
by $3r^{2}\dot{r}$, integrate, and apply the initial conditions to obtain:
\begin{equation}
\left( \frac{dr}{dt} \right)^{2}
=  \frac{2\beta\left(1-r^3\right)}
{3r^3\left( 1 + \epsilon r\right)}.
\label{rfirstintegral}
\end{equation}
A dominant balance is considered to extract the asymptotic collapse behavior near $r=0$, leading to 
\begin{equation}
\frac{dr}{dt}\sim -\left(\frac{2\beta}{3r^3}\right)^{1/2}\text{ as }r\to0,
\label{eq:dombal}
\end{equation}
where the minus sign is chosen to be consistent with bubble collapse.
After integration and application of the constraint $r(1)=0$ that imposes the known dimensional collapse time $T=T_{c2}$ in~(\ref{Tc2}), we obtain
\begin{subequations}
\begin{equation}
r\sim\delta\left(1-t\right)^{2/5} \text{ as } t\to1,
\end{equation}
where
\begin{equation}
\delta=\left(\frac{25\beta}{6}\right)^{1/5}.   
\label{delta}
\end{equation}
\label{asymp}
\end{subequations}
Higher-order (subdominant) corrections to the asymptotic behavior~(\ref{asymp}) are given in Appendix A. A branch point singularity is evident at $t=1$ in both the leading-order and higher-order correction terms. We deduce that this singularity sets the radius of convergence of the power series solution~(\ref{fig:power_series_two_bubble}), and is also responsible for the nonuniform and slow convergence character observed in figure~\ref{fig:power_series_two_bubble} as $t\to1$.

As we have both the (slowly converging) series solution about $t=0$~(\ref{power_series_two_bubble}) and the asymptotic behavior~(\ref{asymp}) as $t\to1$, we apply the method of asymptotic approximants~\cite{AA2017} to develop a more-rapidly converging expression for the bubble collapse.  We assume that the bubble radius may be expressed as an alternative series solution that is consistent with the asymptotic branch point behavior~(\ref{asymp}) as $t\to1$:
\begin{equation}
r = (1-t)^{\frac{2}{5}}\sum_{n=0}^{\infty} A_{n} t^n.
\label{approximant_two_bubble}
\end{equation}
Note that (\ref{approximant_two_bubble}) is an exact solution to~(\ref{rtwobubbles}) in the same way that the original power series solution~(\ref{power_series_two_bubble}) is an exact solution; it is merely a re-summation of the same series, where the portion containing the branch point singularity (responsible for slow convergence) has been infinitely summed and factored out front.  In doing so, the resulting infinite series embedded in~(\ref{approximant_two_bubble}) is expected to converge more rapidly than the original series solution~(\ref{power_series_two_bubble}).  Note that a similar form to~(\ref{approximant_two_bubble}) is proposed in~\cite{OBF2012} with a $(1-t^2)^{2/5}$ replacing the $(1-t)^{2/5}$ in equation~(\ref{approximant_two_bubble}) to impose the evenness in the solution term by term.  This form is consistent with what we formally motivated via the dominant balance~(\ref{asymp}), once one recognizes that $(1-t^2)^{2/5}\sim2^{2/5}(1-t)^{2/5}$ as $t\to1$.  Whereas the form in~\cite{OBF2012} imposes evenness with the alternative prefactor, our form~(\ref{approximant_two_bubble}) is even by construction via the infinite sum.

The coefficients $A_n$ of~(\ref{approximant_two_bubble}) are determined such that the power series expansion 
of~(\ref{approximant_two_bubble}) about $t=0$ matches the (explicitly even) power series solution of~(\ref{rtwobubbles}) centered around $t=0$ given by~(\ref{power_series_two_bubble}).  Therefore, to compute the $A_{n}$ coefficients, we enforce
\begin{equation*}
\sum_{n=0}^{\infty} A_{n} t^n = 
\left( 1 - t \right)^{-\frac{2}{5}}
\sum_{n=0}^{\infty} a_{n} t^{n}.
\end{equation*}
Since the binomial expansion gives 
\begin{equation*}
\left( 1 - t \right)^{-\frac{2}{5}} = \sum_{k=0}^{\infty} {\frac{2}{5}+k-1 \choose k} t^{k}=\sum_{k=0}^{\infty} \frac{\Gamma(k+2/5)}{\Gamma(2/5)\Gamma(1+k)} t^{k}
\end{equation*}
then 
\begin{equation*}
A_{n} = 
\sum_{j=0}^{n} a_{n-j} \frac{\Gamma(j+2/5)}{\Gamma(2/5)\Gamma(1+j)}.
\end{equation*}
The approximant based on~(\ref{approximant_two_bubble}) is written as 
\begin{subequations}
\begin{equation}
r_N = (1-t)^{\frac{2}{5}}\sum_{n=0}^{N+1} A_{n} t^n,
\end{equation}
where
\begin{equation}
A_{n\le N} = 
\sum_{j=0}^{n} a_{n-j} \frac{\Gamma(j+2/5)}{\Gamma(2/5)\Gamma(1+j)}
\text{ and }
A_{N+1}=\delta-\sum_{n=0}^{N} A_{n}.
\end{equation}
\label{approximant}
\end{subequations}
This enforces that the expansion of~(\ref{approximant}) about $t=0$ is exactly~(\ref{power_series_two_bubble}) up to $N$ terms and both the asymptotic behavior and constant $\delta$ is matched as $t\to1$. 

\section{Results and Discussion}
\begin{figure}[h!]
\begin{tabular}{c}
(a)
\includegraphics[width=3in]{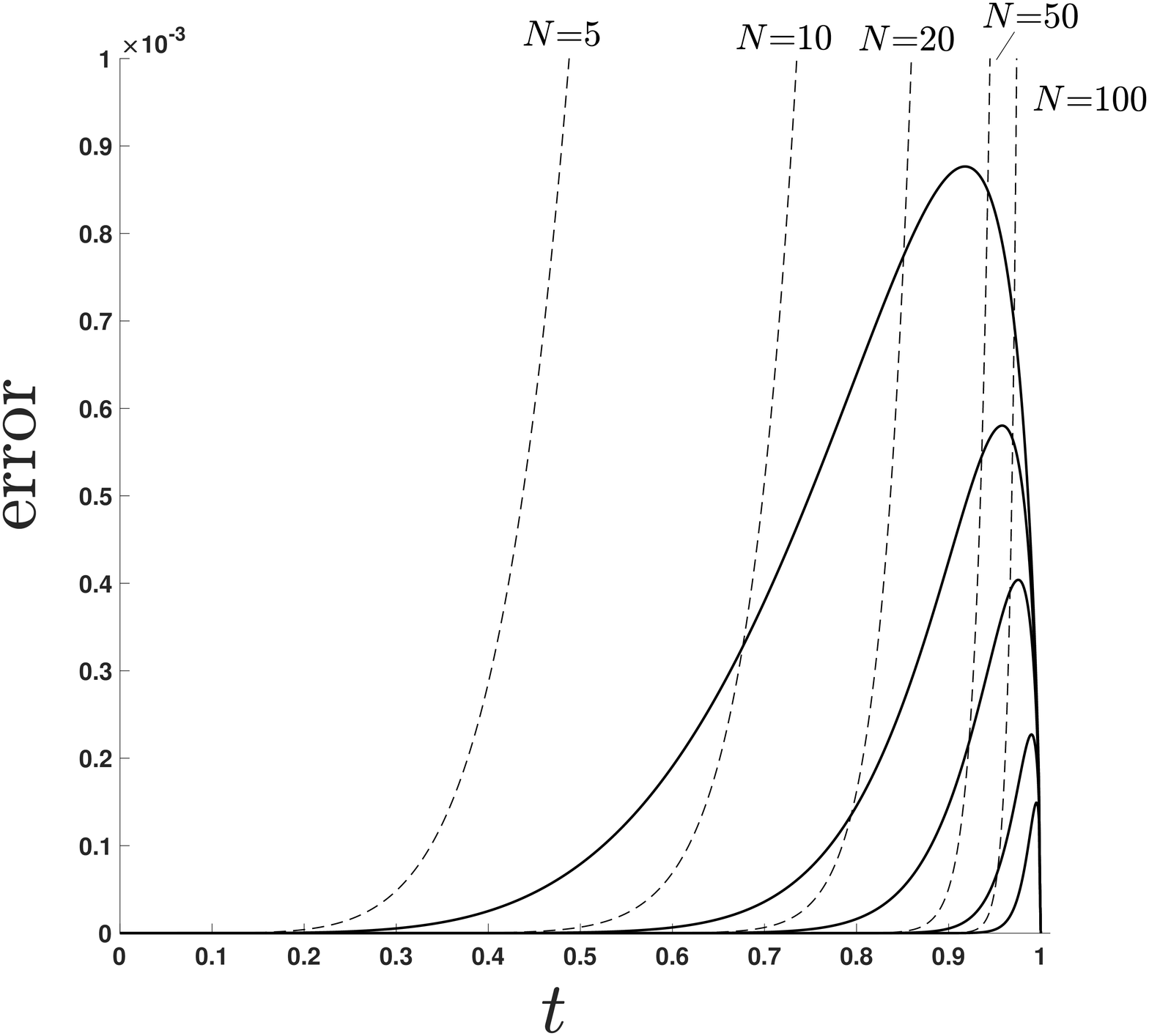}
(b)
\includegraphics[width=3in]{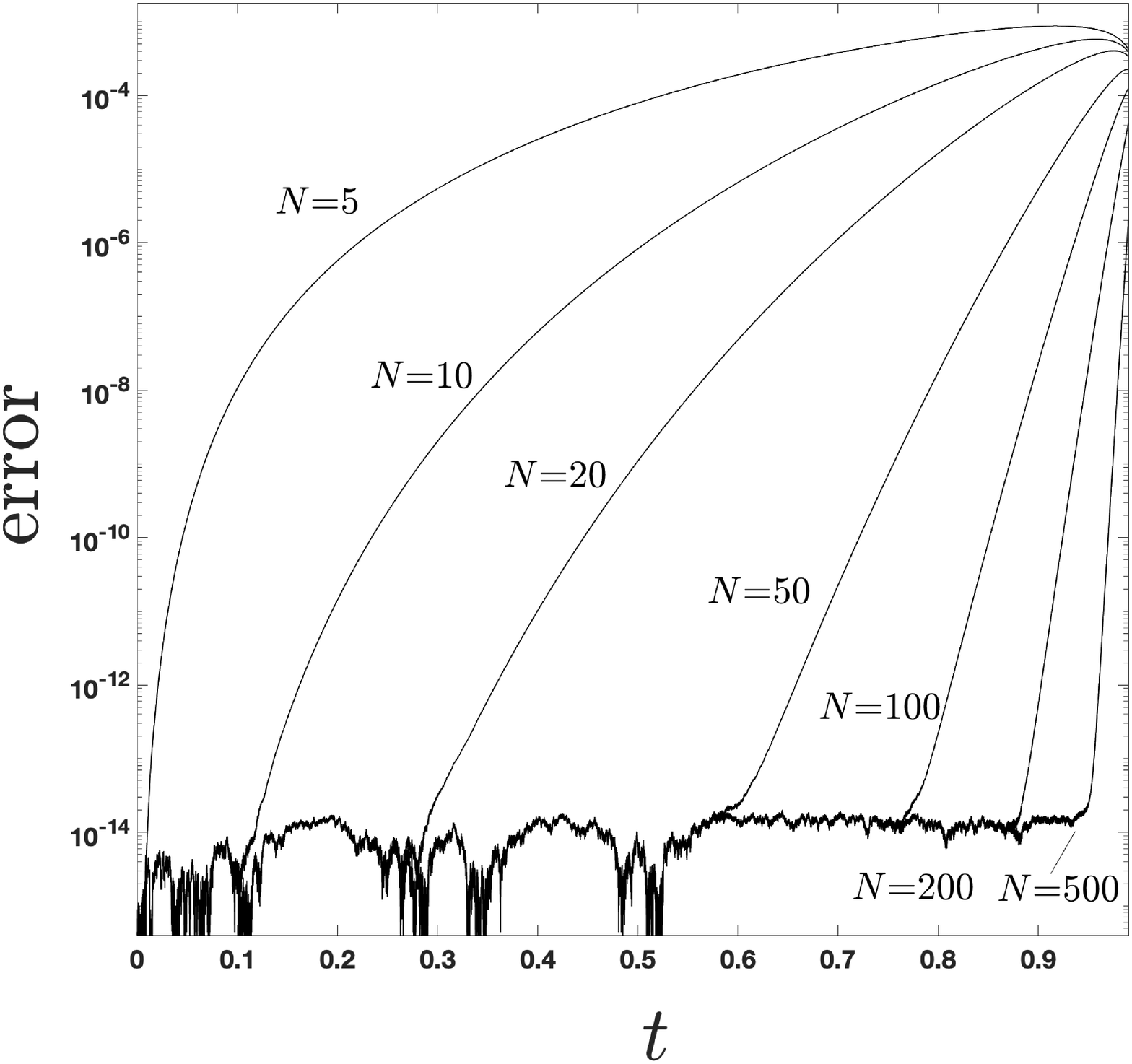}
\end{tabular}
\vspace*{-0.2in}
\caption{Convergence of the approximant~(\ref{approximant}) compared with that of the power series solution~(\ref{power_series_two_bubble}) for $\epsilon=0.1$.  (a) The errors in the $N$-term partial sums of the power series solution~(\ref{power_series_two_bubble}) (dashed curves) are compared with the $N$-term approximant (solid curves) given by~(\ref{approximant}).  The $N$ values for the solid curves correspond to those of the dashed curves from top to bottom in the figure.  The error is the magnitude of the difference between the $N$-term series or approximant and the numerical solution to~(\ref{rtwobubbles}).  (b) The error from part (a) for the $N$-term approximants is provided on a logarithmic scale.}
\label{fig:error_two_bubble}
\end{figure}

We now examine the ability of the approximant~(\ref{approximant}) to provide an analytical representation of the exact solution (taken here to be the numerical solution).  We note that for $N>1$, the approximant is visually indistinguishable from the numerical curve on the scale of Figure~\ref{fig:power_series_two_bubble} for all $\epsilon$ values surveyed ($0\le\epsilon\le0.5$). Thus, in what follows, we provide error plots to examine the ability of the approximant to match the numerical result. Figure~\ref{fig:error_two_bubble}a compares the accuracy of the $N$-term partial sums 
of the series expansion~(\ref{power_series_two_bubble}) (dashed lines) and the $N$-term approximant~(\ref{approximant}) (solid lines) with the numerical solution 
of the two-bubble Rayleigh equation (\ref{rtwobubbles}).  Figure~\ref{fig:error_two_bubble}a shows that the approximant converges with increasing $N$ over the entire physical domain ($0\le t\le1$).  As expected from the convergence trends observed in Figure~\ref{fig:power_series_two_bubble}, the nonuniform convergence behavior of the power series in $t$ is clearly seen in Figure \ref{fig:error_two_bubble}a, 
with additional terms required to achieve accuracy in the vicinity of $t=1$. By comparison, note that even at $N=5$, the approximant shows a maximum error of $O(10^{-3})$, and this error decreases with increasing $N$.  Thus, even with a small number of terms, the approximant provides a favorable representation of the solution, and desired accuracy in practice may be achieved by increasing $N$.  %For example, at 100 terms, the max error is aproximately reduced to approximately 10^-4.

Figure~\ref{fig:error_two_bubble}b shows convergence of the approximant to the numerical 
solution on a logarithmic scale, indicating near-machine-precision accuracy for much of the 
domain.  Figure~\ref{fig:error_two_bubble}b also reveals that, despite improvement of the 
approximant over the power series solution, convergence of the approximant is relatively 
slow near $t=1$ compared with other values of $t$. The origin of this relatively slow 
convergence is higher-order branch point singularities at $t=1$ (see Appendix A); these 
continue to impose the radius of convergence at $t=1$ (from the original 
expansion~(\ref{power_series_two_bubble})) on the $A_n$ series embedded in 
approximant~(\ref{approximant}).  Note also that typical algorithmic tests of convergence, 
such as the well-known ratio test, are inconclusive at the radius of convergence. Thus a 
specific examination of series convergence is required at its radius. In general, when a 
power series converges at its radius of convergence, its convergence will generally be 
slower than elsewhere within its convergence domain.  This behavior is evident in 
Figure~\ref{fig:error_two_bubble} by inspection.
 
\begin{figure}[h!]
\includegraphics[width=4.25in]{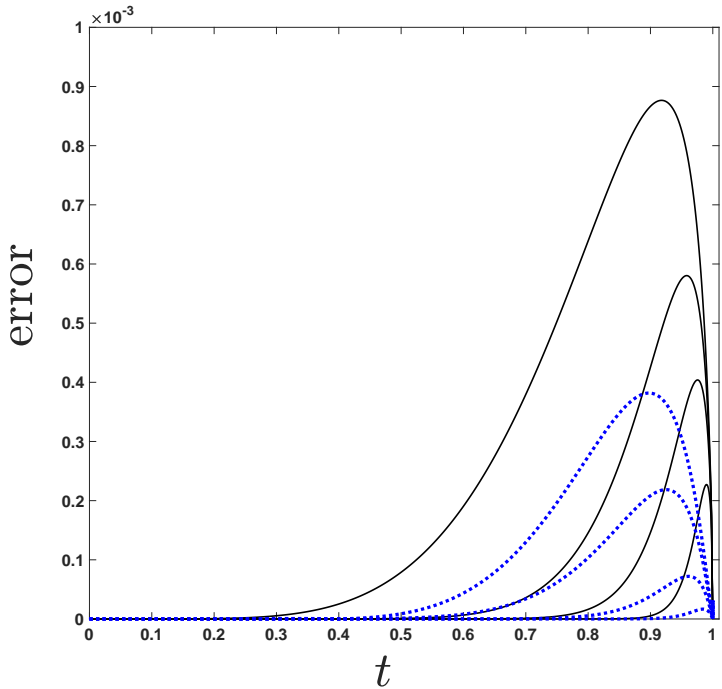}
\vspace*{-0.2in}
\caption{The errors in the $N$-term approximant given by~(\ref{approximant}) for 
$\epsilon$=0.1 (solid curves) and $\epsilon$=0 (thick dashed curves). 
Top to bottom: $N$=5, 10, 20, 50.  The error is the magnitude of the difference between 
the corresponding $N$-term partial sum and the numerical solution to~(\ref{rtwobubbles}).}
\label{fig:error_epsilon}
\end{figure}

Figure~\ref{fig:error_epsilon} provides a comparison of the convergence properties for the 
single ($\epsilon=0$) and two-bubble ($\epsilon\neq0$) cases.  It is apparent that the single 
bubble approximant converges more rapidly near $t=1$ than that of the two-bubble approximant
for a given value of $N$.  This trend continues with larger values of $\epsilon$ (not shown) 
\--- larger $\epsilon$ require larger values of $N$ in the approximant to achieve the same 
accuracy as with smaller $\epsilon$ values. The origin of this different convergence behavior 
again resides with the higher-order corrections shown in Appendix A.  When $\epsilon=0$, the 
higher order correction is $O((1-t)^{8/5})$, while when $\epsilon\neq0$ the correction is $O((1-t)^{4/5})$.  The singularity is thus more severe when $\epsilon\neq0$, and its impact 
on convergence is apparent.   Although one could attempt to create an approximant that captures 
this additional behavior, the simplicity of~(\ref{asymp}) and its accuracy was deemed sufficient 
for applications, and the advantage to handle both the single ($\epsilon$=0) and two-bubble
($\epsilon\neq0$) cases with a single approximant form was deemed attractive.  An example of 
an asymptotic approximant that includes higher order behaviors to improve accuracy is given
in~\cite{BH2018}.

\section{Conclusions}

In this work, exact analytic solutions are provided for the nonlinear ordinary differential equation governing the collapse of two spherical cavities of arbitrary separation distance.  In the limit as separation becomes infinite, the solution reduces to the Rayleigh collapse of a single bubble.  In practice, any infinite series must be truncated and so the rate of convergence is of paramount importance to the utility of the solution form. In this paper we have seen that the standard power series solution (about zero time) to the Rayleigh equation converges slowly on approach to zero radius.  Thus, we utilize the method of asymptotic approximants to enforce the correct asymptotic behavior on approach to collapse in a resummation of the original series. In this application, the branch point singularity at the collapse time imposes a radius of convergence on the original power series.  The resummation explicitly removes the branch point behavior from the infinite sum used, thus making the resulting series behavior more rapidly convergent. Again the approximate is constructed so that the behavior near $t=0$ is matched precisely. There are residual higher-order branch point singularities that continue to limit the convergence of the approximant near collapse time, but their effect is significantly weaker than that of the leading-order singularity.  As in previous applications of the method~\cite{BSWK2012,BSWK2015,BH2018,AA2017,FS2020,SIR2020,SEIR2020}, the results demonstrate that asymptotic approximants not only accelerate convergence, but also (by construction) provide accurate 
analytical expressions that are capable of linking known limiting behaviors in disparate regions of a domain.

\section*{Acknowledgement}
The authors would like to thank Elizabeth Dussan V. for insightful conversations during the course of this work.

\appendix
\section{Higher-Order Corrections For Bubble Collapse for Small Radius}
The leading order behavior for bubble collapse as $r\to0$, is given by~(\ref{asymp}), which is obtained by the method of dominant balance and application of the constraint that $r=0$ at $t=1$.  Here, we determine the next order correction for the bubble collapse as $t\to1$; this corresponds to the limit $r\to0$ by construction.  To begin, we substitute definition~(\ref{delta}) into equation~(\ref{eq:dombal}) and rearrange to yield
\begin{equation}
  {{r}^{3}}{{\left( \frac{dr}{dt} \right)}^{2}}=\frac{4{{\delta }^{5}}(1-{{r}^{3}})}{25(1+\epsilon r)}.
  \label{A.1}
\end{equation}
We assume:
\begin{subequations}
\begin{equation}
    r\sim\delta {{(1-t)}^{2/5}}+D(t),\text{ as }t\to 1,	
    \label{A.2a}
\end{equation}
where $D(t)$ is the as of yet undetermined next-order correction that satisfies
\begin{equation}
D(t)<<{{(1-t)}^{2/5}},~\frac{dD}{dt}\ll {{\left( 1-t \right)}^{-3/5}}\text{ as }t\to 1.	
\label{A.2b}
\end{equation}
\label{A.2}
\end{subequations}
We substitute asymptotic relations~(\ref{A.2}) into~(\ref{A.1}) and employ the method of dominant balance. The following expressions enable simplification of~(\ref{A.1}):
\begin{subequations}
	\begin{equation}
	    	{{\left( \frac{dr}{dt} \right)}^{2}}\sim\frac{4}{25}{{\delta }^{5}}{{\left( 1-t \right)}^{-6/5}}-\frac{4}{5}\delta {{\left( 1-t \right)}^{-3/5}}\frac{dD}{dt}+O\left( {{\left( \frac{dD}{dt} \right)}^{2}} \right)\text{ as }t\to 1	
	    	\label{A.3a}
	\end{equation}
\begin{equation}
    	{{r}^{3}}\sim{{\delta }^{3}}{{\left( 1-t \right)}^{6/5}}+3{{\delta }^{2}}D{{\left( 1-t \right)}^{4/5}}+O\left( {{D}^{2}}{{(1-t)}^{2/5}} \right)\text{ as }t\to 1
    	\label{A.3b}
\end{equation}
\begin{equation}
    {{r}^{3}}{{\left( \frac{dr}{dt} \right)}^{2}}\sim\frac{4}{25}{{\delta }^{5}}-\frac{4}{5}{{\delta }^{4}}{{\left( 1-t \right)}^{3/5}}\frac{dD}{dt}+\frac{12}{25}{{\delta }^{4}}D{{\left( 1-t \right)}^{-2/5}}+O\left( D\frac{dD}{dt}{{\left( 1-t \right)}^{1/5}} \right)\text{ as }t\to 1	\label{A.3c}
\end{equation}
\begin{equation}
    	{{(1+\epsilon r)}^{-1}}\sim1-\epsilon \left( \delta {{\left( 1-t \right)}^{2/5}}+D-\epsilon {{\delta }^{2}}{{\left( 1-t \right)}^{4/5}}+O\left( D{{\left( 1-t \right)}^{2/5}} \right) \right)\text{ as }t\to 1	
    	\label{A.3d}
\end{equation}
\begin{align}
    (1-{{r}^{3}}){{(1+\epsilon r)}^{-1}}\sim1-\epsilon \left( \delta {{\left( 1-t \right)}^{2/5}}+D-\epsilon {{\delta }^{2}}{{\left( 1-t \right)}^{4/5}}+O\left( D{{\left( 1-t \right)}^{2/5}} \right) \right)\nonumber\\
	-{{\delta }^{3}}\left( {{\left( 1-t \right)}^{6/5}}+O\left( D{{\left( 1-t \right)}^{4/5}} \right) \right)\text{ as } t\to 1	
	\label{A.3e}
\end{align}
\label{A.3}
\end{subequations}
In~(\ref{A.3}), the terms listed in the big ``$O$'' are the first neglected terms in the series, and are small compared with the retained terms as $t\to 1$ upon use of~(\ref{A.2}).  Each expansion is truncated at appropriate order to determine $D(t)$ and magnitude of first neglected terms.  The relations~(\ref{A.3}) are substituted into equation~(\ref{A.1}), terms are canceled, and the following equation is obtained:
\begin{subequations}
\begin{equation}
\frac{4}{5}{{\left( 1-t \right)}^{3/5}}\frac{dD}{dt}-\frac{12}{25}D{{\left( 1-t \right)}^{-2/5}}\sim\frac{4}{25}{{\delta }^{2}}G(t)\text{ as }t\to 1	
\label{A4a}
\end{equation}
where:
\begin{equation}
    	G(t)\sim{{\delta }^{2}}{{\left( 1-t \right)}^{6/5}}+O\left( D{{\left( 1-t \right)}^{4/5}} \right)\text{ as } t\to 1 \text{ for } \epsilon =0	
    	\label{A4b}
\end{equation}
\begin{equation}
    	G(t)\sim\epsilon {{\left( 1-t \right)}^{2/5}}+O\left( D \right)+O\left( {{\left( 1-t \right)}^{4/5}} \right) \text{ as } t\to 1 \text{ for } \epsilon \ne 0.
    	\label{A4c}
\end{equation}
\label{A.4}
\end{subequations}
The solution of~(\ref{A.4}) is given as:
\begin{subequations}
\begin{equation}
   	D(t)\sim-\frac{1}{11}{{\delta }^{4}}{{\left( 1-t \right)}^{8/5}}\text{ as }t\to 1\text{ for } \epsilon =0	
   	\label{A5a} 
\end{equation}
\begin{equation}
    	D(t)\sim-\frac{1}{7}\epsilon {{\delta }^{2}}{{\left( 1-t \right)}^{4/5}}\text{ as }t\to 1\text{ for }\epsilon \ne 0
    	\label{A5b}
\end{equation}
	\label{A.5}
\end{subequations}
By inspection, the expressions~(\ref{A.5}) are consistent with the neglected terms in~(\ref{A.3}) through~(\ref{A.4}), which verifies the dominant balance.  The completed asymptotic expressions for $r$, including the next order correction, are assembled by substituting~(\ref{A.5}) in~(\ref{A.2}).  By inspection, it is apparent  that the next order correction for the single bubble ($\epsilon =0$) and 
two- bubble ($\epsilon \ne 0$) cases are different. Furthermore, note that the leading order behavior used to construct the approximant in~(\ref{asymp}) does not fully remove the singularities associated with the branch point at $t=1$, affecting the convergence of the (leading-order-based) approximant discussed in Section 4.

\end{document}